\documentclass[11pt]{article}
\usepackage{amssymb, amsmath}
\usepackage{color, pdfcolmk}
\usepackage{graphics}
\usepackage{graphicx}
\usepackage{amssymb}
\usepackage{epsf}
\usepackage{verbatim}
\newtheorem{theorem}{Theorem}

\newtheorem{lemma}[theorem]{Lemma}
\newtheorem{proposition}[theorem]{Proposition}
\newtheorem{corollary}[theorem]{Corollary}

\definecolor{OliveGreen}{cmyk}{0.64,0,0.95,0.40}
\definecolor{Purple}{cmyk}{0.45,0.86,0,0}

\begin{document}

\title{Large attractors in cooperative bi-quadratic Boolean networks. Part I.}

\author{
 \and
German A. Enciso\footnote{Mathematical Biosciences Institute, Ohio State University, and Harvard Medical School, Department of Systems Biology.}
\ and Winfried
Just\footnote{Department of Mathematics, Ohio University.\newline
This material is based upon work supported by the National Science Foundation
under Agreement No. 0112050 and by The Ohio State University.}
}

\maketitle

\vspace{1ex}


\vspace{1ex}

\begin{abstract}
Boolean networks have been the object of much attention, especially since S. Kauffman proposed them in the 1960's as models for gene regulatory networks.   These systems are characterized by being defined on a Boolean state space and by simultaneous updating at discrete time steps. Of particular importance for biological applications are networks in which the indegree for each variable is bounded by a fixed constant, as was stressed by Kauffman in his original papers.

An important question is which conditions on the network topology can rule out
 exponentially long periodic orbits in the system.
In this paper, we consider systems with positive feedback interconnections among all variables
 (known as cooperative systems), which in a continuous setting guarantees a very stable dynamics.  We show that for an arbitrary constant $0<c<2$ and sufficiently large $n$ there exist $n$-dimensional cooperative Boolean networks
in which both the indegree and outdegree of each variable is bounded by two, and which nevertheless contain periodic orbits of length at least $c^n$. In Part II of this paper we will prove an inverse result showing that any system with such a dynamic behavior must in a sense be similar to the example described.
\end{abstract}


\vspace{1ex}



\bigskip


The concept of a \emph{Boolean network} was originally proposed in the late 1960's by Stuart Kauffman to model gene regulatory behavior at the cell level \cite{Kauffman:N1969,Kauffman:JTB1969}.   This type of modeling can sometimes capture the general dynamics of continuous systems in a simplified framework, e.g. without the choice of specific nonlinearities or parameter values; see for instance~\cite{Albert:JTB2003}.   Boolean networks are known and used in several other disciplines such as electrical engineering, computer science, and control theory,
and analogous definitions are known under various names such as sequential dynamical systems
\cite{Laubenbacher:A2001} or Boolean difference equations \cite{Dee:SIAM1984}.

An important class of \emph{continuous} dynamical systems is that of so-called \emph{monotone systems}, which can be roughly characterized by the absence of negative feedback interactions \cite{Sontag:mono, Smith:monotone}.  A special case is that of \emph{cooperative systems}, in which there are no direct inhibitory interactions
 between any two variables.   Monotone and cooperative systems have been used as a modeling tool for gene regulatory systems, for instance in \cite{Angeli:PNAS2004}.  The assumption of monotonicity is a stringent condition which ensures that the system behavior is remarkably stable: for instance, under mild additional assumptions the generic solution of a monotone dynamical system must converge towards an equilibrium.

In the Boolean case, the class of cooperative systems can be described as that corresponding to maps that can be expressed using only AND and OR gates, i.e.\ with no use of negations.  This can be easily seen by considering the disjunctive normal form of the Boolean maps.

An important question in the study of cooperative Boolean networks is whether some of the stability properties of continuous cooperative systems have analogues in the Boolean case.   For instance, does the assumption of cooperativity by itself limit the length of the longest cycle in an $n$-dimensional Boolean system?
It was shown recently through simulations that random Boolean systems tend to have shorter periodic cycles if they are cooperative, or even if they are close to cooperative in the sense of having few negative feedback interactions; see \cite{Sontag:Laub:al:almostmono}, and also \cite{Greil, Tosic}.   Nevertheless, a straighforward use of Sperner's theorem shows that a cooperative $n$ dimensional Boolean system can have a
cycle of length close to $2^n$ for large $n$, see \cite{Gilbert} and more recently \cite{Just:Enciso:embedding, Sontag:mono:Arxiv2007}.

 One would like to know which additional assumptions rule out exponentially long periodic orbits in cooperative
Boolean systems.  In \cite{Just:Enciso:embedding}  suitable adaptations of the notion of \emph{strong cooperativity} \cite{Smith:monotone} to Boolean systems were found that limit the length of periodic orbits to $2^{\sqrt{n \log n}(1 + o(1))}$ or even to $n$, the dimension of the system.
In the present  manuscript we follow up on this question by considering a different class of cooperative Boolean systems
in which both the indegree and the outdegree of the associated digraph is bounded.

We need some definitions.  An $n$-dimensional \emph{Boolean dynamical system} or \emph{Boolean network} is a pair $(\Pi, g)$, where $\Pi = \{0, 1\}^n$ and $g: \Pi \rightarrow
\Pi$.  A state $s(t)$ at time $t$ will be denoted by $s(t) = [s_1(t), \ldots , s_n(t)]$, or simply
 $s = [s_1, \ldots , s_n]$ if time-dependency is ignored.  We will have

\begin{equation}\label{discrete}
s(t+1) = g(s(t)).
\end{equation}

 The \emph{cooperative order } on $\Pi$ is the partial order relation defined by $s \leq r$ iff
 $s_i \leq r_i$ for all $i \in \{1, \ldots , n\}$.  The system is \emph{cooperative} if $s(t) \leq r(t)$ implies
 $s(t+1) \leq r(t+1)$.

 We associate a directed graph $D$ with vertex set $\{1, \ldots , n\}$ with the system.  A pair $<i, j>$ is in the arc set of $D$
 iff there exist states $s, r \in \Pi$ such that $s_i < r_i$ and $s_k = r_k$ for all $k \neq i$ with the property that
 $(g(s_i))_j < (g(r_i))_j$.  We will say that the system is \emph{bi-quadratic} if both the indegree and the outdegree of all vertices
 in $D$ is at most two.

 Already in his 1969 papers \cite{Kauffman:N1969,Kauffman:JTB1969}, Kauffman focused  his attention on Boolean networks where every variable can only be directly affected by a fixed number $K$ of other variables.  In the digraph associated to the network, this corresponds to limiting the indegree of every node to (at most) $K$.
This corresponds to empirical findings about actual gene regulatory networks which show that most genes are directly regulated by a small number of proteins in a scale-free manner \cite{ArnoneDavidson,Tong:Science2004}.  Other studies of biochemical networks show that only very few nodes are involved in the regulation of other chemicals.
Thus large subnetworks of most biochemical networks of interest will also have the property that the outdegree of each node is bounded by a small integer.  Bi-quadratic Boolean networks satisfy both of these restrictions with $K = 2$.  Random Boolean networks with $K = 2$ have been
extensively studied and tend to have dynamics in the \emph{ordered regime,} which is characterized, among other properties, by the absence of
exponentially long attractors (see \cite{origins}
 for a review).  Thus it becomes a natural question whether one
can prove, for cooperative bi-quadratic Boolean networks, a subexponential bound on the length of their periodic orbits, or at least a bound of the form $c^n$ for some constant $c < 2$.  The following theorem shows that this is not the case.

\begin{theorem} \label{theorem counterexample}  Let $c<2$ be arbitrary.  Then for some sufficiently large $n$ there exists an $n$-dimensional, bi-quadratic cooperative Boolean network which contains a periodic orbit of length at least $c^n$.  Moreover the digraph $D$ associated with this network is strongly connected.
\end{theorem}

The last sentence of Theorem~\ref{theorem counterexample} is of interest in connection with the results in \cite{Just:Enciso:embedding}.  There,
we define a local version $D_s$ of $D$ for every state $s$ as follows: A pair $<i, j>$ is in the arc set of $D_s$
 iff there exist a state $r \in \Pi$ such that either $s_i < r_i$ while $s_k = r_k$ for all $k \neq i$, and we have
 $(g(s_i))_j < (g(r_i))_j$, or $r_i < s_i$ while $s_k = r_k$ for all $k \neq i$, and we have
 $(g(r_i))_j < (g(s_i))_j$.  It is shown that if $X$ is a periodic orbit of an $n$-dimensional cooperative Boolean system
 such that $D_s$ is strongly connected for every $s \in X$, then $|X| \leq n$ (Theorem~25 of \cite{Just:Enciso:embedding}).

 The proof of Theorem~\ref{theorem counterexample} uses a construction similar to a small Turing machine operating on a long circular tape.
 In part~II of this paper we will show that if $c$ is sufficiently close to $2$, then all $n$-dimensional bi-quadratic cooperative Boolean
 networks with periodic orbits of length $\geq c^n$ must contain a relatively small subsystem that can be considered a Turing machine operating
 on one or more tapes that retain the values of all other variables.

 The remainder of this note is organized as follows: In Section~\ref{section simple} we introduce the main idea of the construction, but without requiring
 the system to be cooperative and bi-quadratic.
 In Section~\ref{section main} we show how to modify the construction  so that the network will also be cooperative, bi-quadratic and will have a strongly connected
 digraph.

\section{A Simple Counting Model}  \label{section simple}

In this subsection we consider a conceptual model of a (not necessarily bi-quadratic or cooperative) Boolean network with
 periodic orbits of length $2^N$, for arbitrary $N>0$.  We also discuss the problems that are involved in constructing such a network under the restrictions of Theorem~\ref{theorem counterexample}.  Consider the states $s_1,\ldots s_N$, and the system defined by

\begin{equation} \label{simple cycle}
\begin{array}{l}
s_i(t):=s_{i+1}(t-1), \ \ \  i=1,\ldots, N-1,  \\
s_N(t):=\gamma(s_1(t-1),mode(t-1)).
\end{array}
\end{equation}

One can think of $\gamma$ on a conceptual level
 as a Turing machine operating on variables numbered $i = 1, \ldots , N$ whose values are written on a
circular tape.
The variable $mode$ can have one of two possible values for every $t$, namely $mode=\mbox{\it{rotate}}$, and $mode=\mbox{\it{switch}}$, and the function $\gamma$ is defined by

\begin{equation} \label{simple function}
\begin{array}{l}
\gamma(x,\mbox{\it{rotate}})=x,   \\
\gamma(x,\mbox{\it{switch}})=1-x.
\end{array}
\end{equation}

Thus while $mode(t)=\mbox{\it{rotate}}$, iterating this machine will cyclically rotate the values of $s_1,\ldots , s_N$.  Whenever $mode=\mbox{\it{switch}}$, the machine also will rotate the variable values, but it will invert them at the site $s_N$.

Now let us define the value of the variable $mode$, in such a way that this machine behaves like a counter in base two.  Let us require that at the times $t=0,N,2N,3N,\dots$, $mode(t)=\mbox{\it{switch}}$.  For all other times $t$, define

\begin{equation} \label{simple mode}
mode(t):=\left\{ \begin{array}{ll}
mode(t-1), & \mbox{ if } s_1(t-1)=1, \\
\mbox{\it{rotate}}, &  \mbox{ if } s_1(t-1)=0.
\end{array} \right.
\end{equation}

Thus the model turns into \mbox{\it{switch}} mode exactly at the times $t=0,N,2N,\ldots$, and it only returns back to \mbox{\it{rotate}} mode after $s_1(t_1)=0$ for some $t_1>t$.  The following lemma shows in what way this machine is a counter:  if the states of the system encode numbers in binary format appropriately, then $N$ iterations are equivalent to the addition of one unit modulo $2^N$.

\begin{lemma} \label{lemma simple count}
Given any state $s$ of the model, define $\alpha(s):=s_1 2^0 + s_2 2^1 + \ldots + s_N 2^{N-1}$.  Then $\alpha(s(N))=\alpha(s(0))+1$ mod $2^N$.
\end{lemma}

\noindent
\textbf{Proof:}
Consider an initial state $s(0)$ and let $j\geq 0$ be such that $s_i(\eta)=1$, for $1\leq \eta\leq j <N$, and $s_{j+1}(0)=0$.   Note that $\alpha(s(0))<2^N-1$ in this case.
We have $mode(0)=\mbox{\it{switch}}$ by the definition above (\ref{simple mode}).  By (\ref{simple cycle}), $s_1(\eta)=1$ for $0\leq \eta\leq j-1$, $s_1(j)=0$.  Therefore $mode(\eta)=\mbox{\it{switch}}$, for $1\leq \eta\leq j$, and $mode(j+1)=\ldots=mode(N-1)=\mbox{\it{rotate}}$.   At time $t=N$, the variable values have completed a full rotation and returned to their starting points, except that $s_\eta=0$ for $1\leq \eta\leq j$, $s_{j+1}=1$, and $s_{j+2},\ldots s_N$ are unchanged.  Clearly $\alpha(s(N))=\alpha(s(0))+1$ in this case.

It remains to show the result for the case $j=N$, i.e.\  $s_i(0)=1$, for every $i=1,\ldots,N$.  In that case $mode(0)=mode(1)=\ldots mode(N-1)=\mbox{\it{switch}}$ by (\ref{simple cycle}) and (\ref{simple mode}).  In this way every value of the system is inverted at $s_1$ from 1 to 0, so that $s_i(N)=0$ for $i=1\ldots N$.  Therefore $\alpha(s(N))=0=\alpha(0)+1$ mod $2^N$. $\Box$
\bigskip

\begin{corollary} \label{corollary simple count}
 The network given by equations (\ref{simple cycle}), (\ref{simple function}), (\ref{simple mode}), contains a periodic cycle of length at least $2^N$.
\end{corollary}
\noindent
\textbf{Proof:}  Since the variable $mode$ is reset to $\mbox{\it{switch}}$ for $t=0,N,2N,\ldots$,  Lemma~\ref{lemma simple count} applies
 at each of these time points.  Therefore one can start with $s(0)=0$, and apply
Lemma~\ref{lemma simple count} successively to reach states $s(0),s(N),s(2N), \ldots,$ $s((2^N-1)N)$, which are all different from each other.
$\Box$
\bigskip

Importantly, the function $\gamma$ negates the values of the input $x$ in switching mode.   This appears to be an essential non-monotonic component (or negative feedback) of this system.  Nevertheless, it is shown below that in fact one can rewrite our system in such a way that the resulting system is cooperative.

\subsection{A Generalized Counter}

Before proceeding with the proof of the main result, consider the following generalization of the simple counter above.   Instead of individual Boolean values, each variable $s_i$ is now considered to be a vector with $l>1$ Boolean entries, $s_i=(s_i^l,\ldots,s_i^1)$.
We will treat $s_i$ as a binary code for a nonnegative integer $< 2^l$. At each time $t$, the system continues to be in one of two modes $mode(t)=\mbox{\it{switch}}$ or $mode(t)=\mbox{\it{rotate}}$, but the function $\gamma$ is now replaced with a vector function $\Gamma$ which we describe in the next paragraph.

As before, when $mode=\mbox{\it{rotate}}$ we let $\Gamma(x,mode):=x$. When $mode=\mbox{\it{switch}}$, and given $x=(x^l,x^{l-1}\ldots,x^1)\not=(1,\ldots,1)$, let $j$ be such that $x^{\eta}=1$ for $1\leq \eta \leq j<l$, $x^{j+1}=0$.  Define $y$ by letting $y^\eta:=0$    for
$1\leq \eta \leq j$, letting $y^{j+1}:=1$, and $y^\eta:=x^\eta$ for $j+1<\eta\leq l$.  Set $\Gamma(x,\mbox{\it{switch}}):=y$. If $x=(1,\ldots,1)$, set $\Gamma(x,\mbox{\it{switch}}):=(0,\ldots,0)$.   In other words, the function $\Gamma(x,\mbox{\it{switch}})$ is defined as the addition of 1 to the vector $x$, in base 2 and modulo~$2^l$.

We define the generalized system

\begin{equation} \label{vector cycle}
\begin{array}{l}
s_i(t):=s_{i+1}(t-1), \ \ \  i=1,\ldots, N-1,  \\
s_N(t):=\Gamma(s_1(t-1),mode(t-1)),
\end{array}
\end{equation}
where $\Gamma$ is defined as above.  The variable $mode(t)$ has the value $\mbox{\it{switch}}$ for $t=0,N,2N,\ldots$ and for other values of $t$:

\begin{equation} \label{vector mode}
mode(t):=\left\{ \begin{array}{ll}
mode(t-1), & \mbox{ if } s_1(t-1)=(1,\ldots,1), \\
\mbox{\it{rotate}}, &  \mbox{ otherwise. }
\end{array} \right.
\end{equation}

\begin{lemma} \label{lemma vector}  The network defined by equations (\ref{vector cycle}), (\ref{vector mode}) contains a periodic cycle of length at least $2^{N l}$.
\end{lemma}

\noindent
\textbf{Proof:}
For $(x^l,\ldots, x^1)\in \{0,1\}^l$, define $\beta(x):=x^1 2^0  + x^2 2^1 +\ldots + x^l 2^{l-1}$.  Note that $\beta(\Gamma(x,\mbox{\it{switch}}))=\beta(x)+1$ mod $2^l$.   We follow an argument very analogous to Lemma~\ref{lemma simple count} and Corollary~\ref{corollary simple count}.  Let $\alpha(s):=\beta(s_1) (2^l)^0 + \beta(s_2) (2^l)^1 + \ldots + \beta(s_N) (2^l)^{N-1}$.  Thus the vector $(\beta(s_1),\ldots,\beta(s_N))$ can be regarded as the representation of $\alpha(s)$ in base $2^l$.

As in the proof of Lemma~\ref{lemma simple count}, consider an initial state $s(0)$, and let $j\geq 0$ be such that $s_\eta(0)=(1,\ldots,1)$, for $1\leq \eta\leq j <N$, and $s_{j+1}(0)\not=(1,\ldots,1)$.   As before, we have $mode(\eta)=\mbox{\it{switch}}$ for $0\leq \eta\leq j$, and
 $mode(j+1)=\ldots=mode(N-1)=\mbox{\it{rotate}}$.   At time $t=N$ we have $s_\eta=(0,\ldots,0)$ for $1\leq \eta\leq j$,
as well as $\beta(s_{j+1})=\beta(s_{j+1}(0))+1$, and $s_{j+2},\ldots , s_N$
 are unchanged from $t=0$.  Clearly $\alpha(s(N))=\alpha(s(0))+1$.

In the case that $s_i(0)=(1,\ldots,1)$ for every $i=1,\ldots,N$, it follows as before that $mode(0)=mode(1)=\ldots  = mode(N-1)=\mbox{\it{switch}}$.  Therefore $s_i(N)=(0,\ldots,0)$ for $i=1\ldots N$, and $\alpha(s(N))=0$.

Repeating this process for $s(0)\equiv 0$ and $t=N,2N,\ldots,$ as in Corollary~\ref{corollary simple count}, one finds states $s$ of the system such that $\alpha(s)=1,2,\ldots$, and which are therefore pairwise different.
When $s_i=(1,\ldots,1)$ for all $i$, that is, when $\alpha(s(t))=(2^l)^N-1$, this process reverts to $\alpha(s(t+N)) = 0$. $\Box$
\bigskip

\section{A Cooperative Counter}  \label{section main}

In this section we carry out a construction which is analogous to that in Section~\ref{section simple}, but in which the underlying Boolean network is cooperative, bi-quadratic, and has a strongly connected digraph.  We will need to define some auxiliary Boolean networks with designated input and output variables.

Throughout this section let $L>0$ be an arbitrary even number, and consider the set $A:=\{(r_1,\ldots r_L)\in \{0,1\}^L\,|\, s_1+\ldots+s_L=L/2\}$.   Define the special sequences $\mbox{START}=(1,\ldots,1,0,\ldots,0)$, i.e. $L/2$ ones followed by $L/2$ zeros, and similarly $\mbox{ACTIVE}=$ $(0,\ldots,0,1,\ldots,1)$.

\begin{lemma}  \label{lemma transform}
Let $g:A\to A$ be an arbitrary function.  There exists a Boolean network $B$ with input vectors $a=(a_1,\ldots, a_L)$, $d=(d_1,d_2)$, and output vector $c=(c_1,\ldots, c_L)$, such that for some fixed $m>0$ the following equation holds for every $t$ and $a(t)\in A$, regardless of the initial condition of $B$:

\begin{equation} \label{eq B module}
c(t+m):=\left\{ \begin{array}{ll}
a(t),    &  \mbox{ if } d(t)=(0,1),  \\
g(a(t)), &  \mbox{ if } d(t)=(1,0).
\end{array} \right.
\end{equation}
Furthermore, the network $B$ is cooperative, every node of its associated digraph has in- and outdegree of at most 2, and the indegree (outdegree) of every designated input (output) variable is zero.
\end{lemma}

\noindent
\textbf{Proof:}
Define the set $\hat{A}:=A\times\{ (0,1),(1,0)\}$, and the function $G:\hat{A}\to A$ by $G(x,(1,0)):=g(x)$, $G(x,(0,1)):=x$, for arbitrary $x\in A$.  Since $\hat{A}$ is an unordered set, $G$ can be extended to a cooperative function $G:\{0,1\}^{L+2}\to \{0,1\}^L$; see \cite{Just:Enciso:embedding}.   The result will follow from building a Boolean network that computes the function $G$.

Consider a fixed component $G_i:\{0,1\}^{L+2}\to \{0,1\}$ of $G$.  By the cooperativity of this function, one can write it in the normal form $G_i(y_1,\ldots,y_{L+2})=\Psi_1^i(y_1,\ldots,y_{L+2})\vee\ldots \vee \Psi_{k_i}^i(y_1,\ldots,y_{L+2})$, where each $\Psi_j^i$ is the conjunction of a number of variables, i.e.\ $\Psi_j^i(y_1,\ldots,y_{L+2})=y_{\alpha_{1i}}\wedge\ldots \wedge y_{\alpha_{ji}}$.
This suggests a way of computing $G_i$: define Boolean variables $\psi_j^i(t):=\Psi_j^i(y(t-1))$, and then let $G_i(t):=\psi_1^i(t-1) \vee\ldots \vee \psi_{k_i}^i(t-1)$.   Repeating this procedure for all components of $G$ yields a Boolean network which computes $G$ in $m=2$ steps, and which is cooperative and has indegree (outdegree) zero for every input (output).

In order to satisfy the condition that every node have in- and outdegree of at most two, we need to modify this construction by introducing additional variables.  First, note that the outdegree of every input $y_i$ can be very large.  One can define two additional variables which simply copy the value of $y_i(t)$, then four variables that copy the value of the previous two, etc.  This procedure is repeated for each $y_i$ so that at least as many copies of each variable are present as appear in the expressions of all $\psi_j^i$.  A similar cascade can be used to define each $\psi_j^i$ and  $G_i$ so that each indegree is at most two.  If $\psi_i^j=y_{\alpha_1}\wedge y_{\alpha_2} \wedge y_{\alpha_3}$, say, then one can define $z_1(t):=y_{\alpha_1}(t-1)$, $z_2(t):=y_{\alpha_2}(t-1)\wedge y_{\alpha_3}(t-1)$, $\psi_i^j(t):=z_1(t-1)\wedge z_2(t-1)$.
Similarly for longer disjunctions and each $\psi_j^i$ and also similarly for $G_i$, in which case $\wedge$ is replaced by $\vee$ at each step.  This produces a computation of $G_i$ in $m_i$ steps for each $i$.  Finally, after introducing further additional variables at each component $i$ if necessary to compensate for unequal lengths of the expressions for $\psi^i_j$,
the Boolean vector $G(y_1,\ldots, y_{L+2})$ can be computed in exactly $m=\max(m_1,\ldots,m_{L})$ steps.
$\Box$
\bigskip

\paragraph{Remark:}  Without loss of generality, we can assume that for every state variable $s$ in the network $B$, there exists some input variable $d_i$ or $a_i$ and a directed path from this input towards $s$.  This is because if that wasn't the case, one could delete $s$ from the system without altering equation (\ref{eq B module}).   Cooperativity of $G$ implies that
$G(0,\ldots,0)=(0,\ldots,0)$ and $G(1,\ldots,1)=(1,\ldots,1)$ \cite{Just:Enciso:embedding};
therefore each $G_i$ is non-constant and no output variable will be deleted.
Similarly, it will be assumed that for every state variable $s$, there exists an output variable $c_i$ such that there is a directed path from $s$ to $c_i$.

\vspace{2ex}

Lemma~\ref{lemma transform}
 can be used to compute a function $g$ which will be used in a way analogous to $\gamma$ in equation (\ref{simple cycle}).   Similarly, we need to construct a `switch' to determine when to turn the system into \mbox{\it{rotate}} mode, which is provided by Lemma~\ref{lemma switch} below.
Note that Lemma~\ref{lemma transform} cannot be used for this purpose because the desired output depends not only on the current state of the input
$p(s)$ but on the whole history (of unknown length) of the input sequence since the last time when $p(s)$ took the value $START$.

\begin{lemma} \label{lemma switch}
There exists $\mu>0$ and a Boolean network $D$ with input vector $p=(p_1,\ldots, p_L)$, and output vector $q=(q_1,q_2)$, such that the following holds for any initial condition of $D$.   Consider any sequence of inputs $p(0),p(1), \ldots, p(M)$, $M>1$, such that

\noindent
i)  $p(s)\in A$, for $0\leq s\leq M$,

\noindent
ii) $p(0)=\mbox{START}$,
and

\noindent
iii) $p(s)\not=\mbox{START}$, for $0<s\leq M$.

\noindent
Let $j\geq 0$ be such that $p(s)=\mbox{ACTIVE}$ for $1\leq s\leq j$,  $p(j+1)\not=\mbox{ACTIVE}$ (or $p(1)=\ldots =p(M)=\mbox{ACTIVE}$ and $j=M$).
Then
\begin{equation}  \label{eq switch}
q(s)=\left\{ \begin{array}{ll}
(1,0), &  \mu\leq s \leq \mu+j,  \\
(0,1), &  \mu+j < s \leq  \mu+M,
\end{array}  \right.
\end{equation}
Furthermore, the network $B$ is cooperative, every node of its associated digraph has in- and outdegree of at most 2, and the indegree (outdegree) of every designated input (output) variable is zero.
\end{lemma}

\begin{figure}[ht]
\centerline{\includegraphics[width=4.5in]{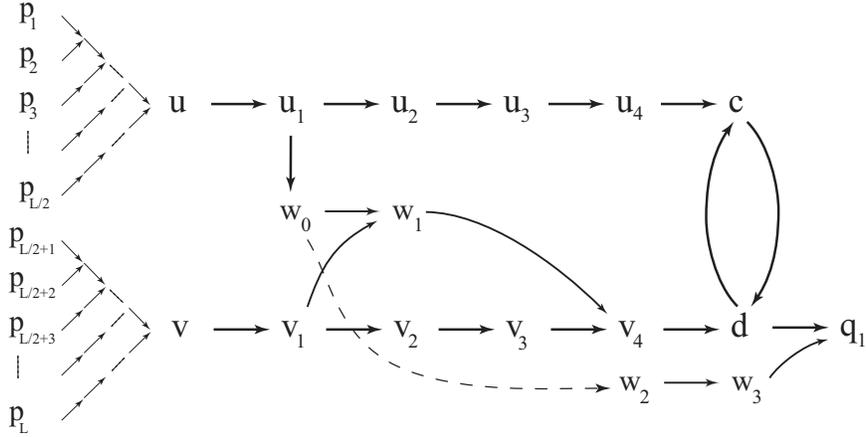} }
\caption{The digraph of the network $D$ which is used to compute the output $q_1$ from the input $p$.  The formulas for each interaction (i.e. $\wedge,\vee$) as well as the dependencies of $u_2$ on $u$ and $v_2$ on $v$ are omitted in this figure.}
\label{figure switch}
\end{figure}

\noindent
\textbf{Proof:}
The idea for this proof is the simple system $c(t)= u(t-1) \vee d(t-1)$, $d(t)= v(t-1) \wedge c(t-1)$, with inputs $u,v$.  This switch is turned on by letting both inputs $u=1$ and $v=1$ for a short time, after which $u$ can be turned to $0$ while $v$ is left equal to~$1$.  After letting $v=0$ for a short time, the switch resets and doesn't restart even if $v=1$ again.

Let $t=0$ without loss of generality, the more general case being completely analogous.   For the sake of clarity assume for now that $0<j<M$,
but the same construction allows for $j=0$ and $j=M$ as described below.  See Figure~\ref{figure switch} which displays the circuit described below.
Define for the moment $u(t):=p_1(t-1) \wedge \ldots \wedge p_{L/2}(t-1)$,
$v(t):=p_{L/2+1}(t-1)\wedge \ldots \wedge p_{L}(t-1)$ (a modification of this definition with additional variables and indegree two is displayed in the
figure and described below).  Thus $u(s)=1$ if and only if $p(s-1)=\mbox{START}$, and $v(s)=1$ if and only if $p(s-1)=\mbox{ACTIVE}$, since by assumption $p(s)\in A$.

Define
\[
u_1(t):=u(t-1),\ u_2(t):=u(t-1)\vee u_1(t-1), \ u_3(t):=u_2(t-1),\ u_4(t):=u_3(t-1),
\]
\[
v_1(t)=v(t-1), \ v_2(t)=v(t-1)\wedge v_1(t-1),\ v_3(t):=v_2(t-1),\ v_4(t):=v_3(t-1)\vee w_1(t-1),
\]
\[
w_0(t):=u_1(t-1),\ w_1(t):=w_0(t-1)\wedge v_1(t-1),
\]
\[
c(t):=u_4(t-1)\vee d(t-1), \ d(t):=v_4(t-1) \wedge c(t-1).
\]

(Intuitively, $u_4$ is a time-transposed copy of $u$ where every 1 has been doubled due to the feed-forward loop at $u_2$.  Also, $v_4$ is similar to a time-transposed copy of $v$ where every 0 has been doubled - the auxiliary variables $w_i$ only play a role at a single time step as described below.  The loop $c \leftrightarrow d$ forms the core of the switch in the system.)

A simple calculation shows that $u_4(4)=u_4(5)=1$, $u_4(s)=0$ for $5<s\leq  M+4$.  On the other hand, since $v(1)=0, v(2)=\ldots=v(1+j)=1, v(2+j)=0$, we infer that
$v_2(2)=v_2(3)=0$, $v_2(s)=1$ for $3<s\leq 2+j$, $v_2(3+j)=v_2(4+j)=0$.   It follows that
$w_1(3)=0$ (since $v_1(2)=0$),
 and that $w_1(4)=1$ if and only if $v_1(3)=1$ (since $w_0(3)=1$).    This in turn holds since $j>0$.  Also, $w_1(s)=0$ for $s>4$.

We use the data for $w_1$ and $v_3$ to compute the values of $v_4$.  From $w_1(3)=v_3(3)=0$, it follows that $v_4(4)=0$.  From $w_1(4)=1$ it follows that $v_4(5)=1$, and using $v_3$ we similarly infer that $v_4(s)=1$ for $s=4< s\leq 4+j$.  Also, $v_4(5+j)=v_4(6+j)=0$.

We conclude that $c(5)=1$, $d(5)=0$, regardless of the  values of $c,d$ at earlier time steps.
  Since $j>0$,
one has $c(6)=1$, $d(6)=1$, and in general $c(s)=d(s)=1$ for $5<s\leq 5+j$.  Then $c(6+j)=1$, $d(6+j)=0$, $c(s)=d(s)=0$, for $7+j\leq s\leq 5+M$, and $d(6+M)=0$.

In particular  $d(s)=1$ for exactly $j$ time steps, $5<s\leq 5+j$, and then  $d(s)=0$ for $6+j\leq s \leq 6+M$.
Since we want the variable $q_1$ to be equal to 1 during exactly $j+1$ time steps, we define the additional variables
\[
w_2(t):=w_0(t-1),\ w_3(t):=w_2(t-1),\ q_1(t):=w_3(t-1)\vee d(t-1).
\]
Calculating that $w_3(5)=1$, $w_3(s)=0$ for $5<s\leq 5+M$, we have $q_1(s)=1$, $6\leq  s\leq 6+j$, and $q_1(s)=0$, $6+j<s\leq 7+M$.

In order to define the variable $q_2$, it suffices to make a construction dual to the previous one (recall that simply negating $q_1$ is not permitted).  That is, define $\hat{u}(t):=p_{L/2+1}(t-1)\vee \ldots \vee p_{L}(t-1)$, and $\hat{v}(t):=p_1(t-1)\vee \ldots \vee p_{L/2}(t-1)$, in such a way that  $\hat{u}(s)=0$ if and only if $p(s-1)=\mbox{START}$, and $\hat{v}(s)=0$ if and only if $p(s-1)=\mbox{ACTIVE}$.   Define variables $\hat{u}_1,\hat{v}_1$ etc.\  similarly as above, except that every $\wedge$ in the function definition is replaced by $\vee$ and vice versa.
Then it will necessarily follow that $q_2= \neg q_1$
 on the interval $6\leq s\leq 6+M$.   Using the value $\mu=6$, equation (\ref{eq switch}) is satisfied.

The case $j=0$ is very similar as above.  In this case $w_1(4)=0$ (instead of 0 for $j>0$), $v_4(4)=v_4(5)=0$, and therefore $d(s)=0$ on all $6\leq s\leq M+6$.  Thus $q_1(6)=1$, and $q_1=0$ for larger values of $s$.  In the case $j=M$, one can compute $v_4(s)=1$ for $5\leq s< M+5$.  This allows the variables $c(s),d(s)$ to remain equal to 1 up to and including $s=M+5$.  Therefore $q_1(s)=1$ up to and including $s=6+M$.

Notice that this system is cooperative, and that all in- and outdegree requirements are satisfied except for the indegree of the variables $u,v,\hat{u},\hat{v}$.   These terms can now be replaced in a routine manner by a cascade of variables (see Figure~\ref{figure switch}),  in such a way that $u(s)=1$ if and only if $p(s-\tau)=\mbox{START}$, etc.,\ for some $\tau>1$.  This will increase the delay $\mu$ but leave the computations and the other properties of this system unchanged.
$\Box$
\bigskip

We are ready for the construction of the cooperative counter described in the introduction.  This Boolean network is designed to replicate the behavior of the system described by equations (\ref{vector cycle}), (\ref{vector mode}), while ensuring its cooperativity.  In order to do so, we let $l>0$ be arbitrary and  $L>0$ be an even positive integer, which is large enough that there exists an injective function $\chi:\{0,1\}^l\to A$, where $A$ is defined as above.  The cooperative network will contain $L$-dimensional vectors $r_i=(r_i^1,\ldots,r_i^L)$, with values in $A$, which will be considered as proxy for states $s_i=\chi^{-1}(r_i)$ of the system (\ref{vector cycle}), (\ref{vector mode}).

We require that $\chi(1,\ldots,1)=\mbox{ACTIVE}$, and that $\mbox{START}\not\in \mbox{Im} (\chi)$ (see the definitions of $\mbox{START}$ and $\mbox{ACTIVE}$ above).  This is possible if $L$ is large enough so that $\binom{L}{L/2}> 2^l$.  We also let $\chi(0,\ldots,0)=(1,0,\ldots,1,0)$, and $\chi(0,0,\ldots,0,1)=(0,1,0,1,\ldots,0,1)$.
Having defined $\chi$, we define $g:A\to A$ as $g(r):=\chi(\Gamma(\chi^{-1}(r)))$, for $r\in \mbox{Im} (\chi)$, $g(r)=r$ for all other $r\in A$. The function $\Gamma$ is defined as in Section~\ref{section simple}.  In particular, $g(\mbox{START})=\mbox{START}$.

Using the function $g$ defined above, we consider the cooperative networks $B$ and $D$ from Lemmas~\ref{lemma transform} and~\ref{lemma switch}.
Recall that $B$ ($D$) has variables $a,d$ ($p$) which are specifically designated as inputs, a variable $c$ ($q$) specifically designated as output, and a  `processing delay' $m$ ($\mu$).  The cooperative network, which will be denoted by $S$, is defined by  $B$ and $D$, together with the equations

\begin{equation} \label{cooperative cycle}
\begin{array}{l}
r_i(t):=r_{i+1}(t-1), \ \ \  i=m+2,m+3,\ldots, N,  \\
r_{N+1}(t):=c(t-1),
\end{array}
\end{equation}
and

\begin{equation} \label{cooperative Turing}
\begin{array}{l}
a(t):=r_{m+2}(t-1),  \\
d(t):=q(t-1),        \\
p(t):=r_{m+\mu+2}(t).
\end{array}
\end{equation}

\begin{figure}[ht]
\centerline{\includegraphics[width=6in]{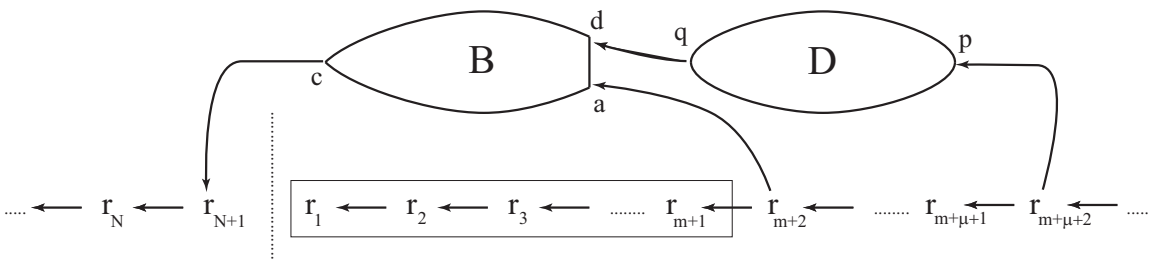} }
\caption{The network interconnections of the system $S$ given by $B$, $D$, and equations (\ref{cooperative cycle}), (\ref{cooperative Turing}).  The variables $r_1,\ldots, r_{m+1}$ are displayed in a box to indicate that they are not part of $S$ but only included in the proof of Theorem~\ref{teo cooperative cycle}.}
\label{figure cooperative}
\end{figure}

See Figure~\ref{figure cooperative} for an illustration.  Since both of the subnetworks used in the construction of this system contain only the Boolean operators $\wedge,\vee$ in their expression (and no negations), it follows from (\ref{cooperative cycle}) and (\ref{cooperative Turing}) that the same is the case for the full network, hence the system is cooperative.

\begin{proposition}   \label{teo cooperative properties}
The digraph of the Boolean network $S$ is strongly connected and bi-quadratic.
\end{proposition}

\noindent
\textbf{Proof:}
The fact that every in- and outdegree is at most 2
 follows directly from equations (\ref{cooperative cycle}), (\ref{cooperative Turing}) and Lemmas~\ref{lemma transform} and~\ref{lemma switch}, taking into account that the indegree (outdegree) of every input (output) variable is zero within their respective subnetwork.
See also Figure~\ref{figure cooperative}.

In order to show the strong connectivity of the digraph, first we show that there exists a directed path from every node in the network to the node $q_1$, the first component in the output of $D$.  It is clear from the circuit defining $D$ that every input variable $p_i$ has a path connecting to $q_1$ (the first $L/2$ components of $p$ through the variables $u,u_1,\ldots$ and the last $L/2$ components through $v,v_1,\ldots$).  Therefore every component of every variable $r_i$ can reach $q_1$ as well.  By the remark after Lemma~\ref{lemma transform}, the same applies to every variable of $c$, and thus to every variable in the subnetwork $B$.
Thus the same applies also to $q_2$, and hence to every state in the subnetwork $D$.

Now we show that there exists a path from $q_1$ to every node in the network.  Suppose first that there exists $c_j$ such that neither $d_1$ or $d_2$ contains a path towards $c_j$.
This would imply that $g_j(x)=x_j$ for every argument $x\in A$, by equation (\ref{eq B module}).  But we have
\[
g(1,0,\ldots,1,0)=\chi(\Gamma(0,\ldots,0))=\chi(0,\ldots,0,1)=(0,1,0,1,\ldots,0,1),
\]
which is a contradiction.  Thus for every $j$, there exists a path from either $d_1$ or $d_2$ to $c_j$ (and therefore from $q_1$ or $q_2$ to $c_j$).

Since there exists a path from $q_1$ to $q_2$, it follows that there is a path from $q_1$ to every~$c_j$.  Thus every component of every state $r_i$, $p$, and $a$ can be reached by a path from $q_1$.  Every state in $B$ can be reached from $d_1$ and hence $q_1$, once again by the remark after Lemma~\ref{lemma transform}; the same applies to $q_2$, and every state in the subnetwork $D$.
$\Box$
\bigskip

\begin{theorem}  \label{teo cooperative cycle}
Let $L>0$ be an even number such that $\binom{L}{L/2}> 2^l$.  Then system $S$ has a periodic orbit of length greater than or equal to $2^{l N}$.
\end{theorem}

\noindent
\textbf{Proof:}
For the purposes of this proof, we extend the system with the auxiliary variables $r_1,\ldots, r_{m+1}$, defined by $r_i(t):=r_{i+1}(t-1), \ \ \  i=1,\ldots, m+1$; see Figure~\ref{figure cooperative}.  These variables cannot change the length of the original system's periodic orbits
 (since they don't feed back into it), but they can nevertheless be used for the study of the network.

Suppose that the system is initiated at time $t_0$, and let $t\geq t_0+m+2$.  Then $r_1(t-1)=r_{m+2}(t-m-2)$ by (\ref{cooperative cycle}).  But then $a(t-m-1)=r_{m+2}(t-m-2)=r_1(t-1)$, by (\ref{cooperative Turing}).  By Lemma~\ref{lemma transform}, $c(t-1)$ is equal to either $r_1(t-1)$ or $g(r_1(t-1))$, depending on whether $d(t-m-1)=(0,1)$ or $(1,0)$ respectively.  Since $r_{N+1}(t)=c(t-1)$, we have

\begin{equation} \label{cooperative mode}
r_{N+1}(t):=\left\{ \begin{array}{ll}
r_1(t-1), & \mbox{ if } mode(t-1)=\mbox{\it{rotate}}, \\
g(r_1(t-1)), &  \mbox{ if } mode(t-1)=\mbox{\it{switch}},
\end{array} \right.
\end{equation}
where the auxiliary Boolean variable $mode(t)$ is defined as $mode(t):=\mbox{\it{switch}}$ if $d(t-m)=(0,1)$ and $mode(t):=\mbox{\it{rotate}}$ if $d(t-m)=(1,0)$.  The variable $mode$, similarly as $r_1,\ldots,r_{m+1}$, is defined merely for the purposes of this proof, and it does not
form part of the network itself.

Suppose now that $t_0\leq -m-\mu-2$.  At time $0$, assume that $r_{N+1}=\mbox{START}$, and $r_\eta\not=\mbox{START}$, for $1\leq \eta\leq N$.   Let $j\geq 0$ be such that $r_\eta=\mbox{ACTIVE}$ for $1\leq \eta\leq j<N$, and $r_{j+1}\not=\mbox{ACTIVE}$.  We show that

\begin{equation} \label{eq mode cooperative}
mode(\eta)=\mbox{\it{switch}},\ \ 0\leq \eta\leq j;\ \ mode(\eta)=\mbox{\it{rotate}},\ \ j+1\leq \eta\leq N.
\end{equation}

To see this, note that by (\ref{cooperative cycle}) $r_{m+\mu+2}(\eta)\not=\mbox{START}$, for $-m-\mu-1\leq \eta\leq N-m-\mu-2$.  Since $g^{-1}(\mbox{START})=\mbox{START}$, it also follows that $\mbox{START}=r_1(0)=c(-1)=a(-m-1)=r_{m+2}(-m-2)$, and $r_{m+\mu+2}(-m-\mu-2)=\mbox{START}$.  Thus
setting $t=-m-\mu-1$, one has $p(t)=\mbox{START}$, $p(\eta)\not=\mbox{START}$ for $t<\eta\leq t+N$, $p(\eta)=\mbox{ACTIVE}$ for $t+1\leq\eta\leq t+j$, and $p(t+j+1)\not=\mbox{ACTIVE}$.  Applying Lemma~\ref{lemma switch} with $M:=N$, we have that $q(\eta)=(1,0)$, for $t+\mu\leq \eta\leq t+\mu+j$, and $q(\eta)=(0,1)$, $t+\mu+j<\eta\leq t+\mu+N$.  Equation~(\ref{eq mode cooperative}) then follows directly from the definition of $d$ and the mode variable.   It is analogous to verify that (\ref{eq mode cooperative}) also holds in the case $j=N$, i.e.\ when $r_{\eta}(0)=\mbox{ACTIVE}$ for $1\leq \eta\leq N$.

Note that using equation (\ref{eq mode cooperative}) we can fully calculate $r(N+1)$, namely $r_\eta(N+1)=g(r_\eta(0))$, for $1\leq \eta\leq j+1$, and $r_\eta(N+1)=r_\eta(0)$ for $j+1<\eta\leq N$; also, necessarily $r_{N+1}(N+1)=\mbox{START}$ regardless of $mode(N)$, since $g(\mbox{START})=\mbox{START}$.
The same process can be repeated starting at time $N+1$, $2(N+1),$ etc., since necessarily $r_\eta=\mbox{START}$ can still only hold for $\eta=N+1$.

An appropriate initial condition to reach the above situation can be given as follows.  Let $t_0=-(N+1)$, and let $r_\eta(t_0)=\chi^{-1}(0,\ldots,0)$, for $1\leq \eta\leq N$.  Let $r_{N+1}(t_0)=\mbox{START}$.   Finally, let $B$ ($D$) be initialized with $m$ ($\mu$) successive inputs of $a=\chi^{-1}(0,\ldots,0),\ d=(0,1)$  ($p=\chi^{-1}(0,\ldots,0)$).   This way for $t=0$ we guarantee that $r_{N+1}=\mbox{START}$, $r_\eta(t_0)=\chi^{-1}(0,\ldots,0)$ for $1\leq \eta\leq N$, and importantly, $t_0\leq -m-\mu-2$.

Finally, under our standing hypotheses $t_0\leq -m-\mu-2$, $r_{N+1}(0)=\mbox{START}$, and $r_\eta(0)\not=\mbox{START}$ for $1\leq \eta\leq N$. Define
 the following initial conditions for the system~(\ref{vector cycle}),~(\ref{vector mode}):  $s_\eta(0):=\chi^{-1}(r_\eta(0))$, $i=1\ldots N$.  After calculating $j$ as before, $0\leq j\leq N$, we have seen that $s_\eta(N)=\Gamma(s_\eta(0))$ for $1\leq \eta\leq j+1$, and $s_\eta(N)=s_\eta(0)$ otherwise.  From the discussion above, it follows that $\chi^{-1}(r_\eta(N+1))=s_\eta(N)$ for $1\leq \eta\leq N$.  This equivalence between the two systems implies in particular that the states $r(t)$ are pairwise different for $t=0,N+1,2(N+1),\ldots,(2^{l N}-1)(N+1)$.  The result follows.
$\Box$
\bigskip

\subsection{Proof of Theorem~\ref{theorem counterexample}}

We can use Proposition~\ref{teo cooperative properties} and Theorem~\ref{teo cooperative cycle} to prove the theorem stated in the introduction.  Let $0< c<2$ be arbitrary.  We prove first that there exist $L>0$ even and and integer $l>0$ such that

\begin{equation} \label{L l condition}
\binom{L}{L/2} > 2^l > c^L.
\end{equation}

The second inequality is equivalent to $L/l < \ln 2/ \ln c$; thus let $L=w l$, for some fixed $1<w<\ln 2/ \ln c$ (for large enough $l$, $L$ can then be rounded up to the nearest even number while satisfying this inequality).  Using Stirling's formula, we have $\binom{L}{L/2} > v\, 2^L/\sqrt{2\pi L}$ for large enough $L$, where $0<v<1$ is arbitrary and fixed.  The first inequality in (\ref{L l condition}) is satisfied if $v\, 2^L/\sqrt{2\pi L}>2^l$. But after replacing $L=w l$ this is equivalent to $2^{(w-1)l}> v^{-1}\sqrt{2\pi w l}$.  Clearly this inequality is satisfied for sufficiently large $l$, hence (\ref{L l condition}) follows.

The first inequality is now used to carry out the construction of system $S$, which by Theorems~\ref{teo cooperative properties} and~\ref{teo cooperative cycle} is cooperative and bi-quadratic with strongly connected digraph, and has a
 periodic orbit of length greater than or equal to $2^{N l}$.

 It remains to show that $2^{N l} \geq c^n$ for large $N>0$, where $n$ is the dimension of the system.
Let $T$ be the total
number of variables in the subnetworks $D,B$.  Note that $T$ depends only on $L,l$, and not on $N$.  Then $n=(N+1-(m+1))L+T=NL-mL+T$.  Notice that $c^n\leq 2^{N l}$ if and only if $(NL-mL+T)\ln c \leq N l \ln 2$, which holds if and only if

\[
L \ln c \leq l \ln 2 + \frac{mL-T}{N} \ln c.
\]
But this equation is satisfied for large enough $N$, since $L \ln c< l \ln 2$ by (\ref{L l condition}). $\Box$

\end{document}